\documentclass[pra,twocolumn,aps,showpacs,floatfix,superscriptaddress]{revtex4-1}

\usepackage{graphicx}
\usepackage{rotating}
\usepackage{amsmath}
\usepackage{bbm}
\usepackage{subfigure}
\usepackage{color}

\renewcommand{\v}[1]{{\bf #1}}

\newcommand{\be}{\begin{equation}}
\newcommand{\ee}{\end{equation}}
\newcommand{\bea}{\begin{eqnarray}}
\newcommand{\eea}{\end{eqnarray}}
\newcommand{{\br}}{\bf r}

\graphicspath{
  {./}
  {figures/pdf/}
  {figures/eps/}
  {figures/jpg/}
}

\begin{document}

\title{ Quasi-particle energy spectra in local reduced density matrix functional theory}
\author{Nektarios~N.\ Lathiotakis}
\affiliation{Theoretical and Physical Chemistry Institute, National Hellenic 
Research Foundation, Vass.  Constantinou 48, GR-11635 Athens, Greece}
\author{Nicole\ Helbig}
\affiliation{Peter-Gr\"unberg Institut and Institute for Advanced Simulation,
Forschungszentrum J\"ulich, D-52425 J\"ulich, Germany}
\author{Angel~Rubio}
\affiliation{Nano-Bio Spectroscopy
group and ETSF Scientific Development Centre, Dpto.\ F\'isica de Materiales,
Universidad del Pa\'is Vasco, CFM CSIC-UPV/EHU-MPC and DIPC, Av.\ Tolosa 72,
E-20018 San Sebasti\'an, Spain}
\author{Nikitas~I.~Gidopoulos}
\affiliation{Department of Physics, Durham University, South Road,  
Durham DH1 3LE, United Kingdom}

\begin{abstract}  
\noindent
Recently, we introduced (Phys. Rev. A, {\bf 90}, 032511 (2014)) {\em local reduced density
matrix functional theory} (local RDMFT), a theoretical scheme capable of incorporating static correlation effects 
in Kohn-Sham equations.
Here, we apply local RDMFT to  molecular systems of relatively large size, as a demonstration of its computational 
efficiency and its accuracy in predicting single-electron properties from the eigenvalue spectrum of the 
single-particle Hamiltonian with a local effective potential.
We present encouraging results on the photoelectron spectrum of 
molecular systems and the relative stability of C$_{20}$ isotopes.
In addition, we propose a modelling of the fractional occupancies as functions of the orbital energies that further 
improves the efficiency of the method useful in applications to large systems and solids.
\end{abstract}

\date{\today}

\maketitle


\section{introduction\label{sec:intro}}

In electronic structure theory, a desirable and elegant feature of independent particle models, such as the Hartree-Fock equations or the Kohn-Sham scheme, is the direct 
prediction of single-electron properties, like ionization potentials (IPs), from the eigenvalue spectrum of their corresponding effective single-particle Hamiltonians. 
For example, in Hartree-Fock (HF) theory, Koopmans showed~\cite{K1934} that the eigenenergies, $\epsilon_i$, 
of the occupied molecular orbitals are equal to minus the corresponding ionization potentials, $I_i=-\epsilon_i$, 
within the approximation that the other occupied orbitals remain frozen.
%
Also, in (exact) Kohn-Sham (KS) density functional theory (DFT)
the energy of the highest occupied molecular orbital (HOMO) equals the first ionization potential of the system~\cite{PPLB1982}. 
Further, by inverting accurate ground state densities to obtain a good approximation of the exact KS potential, 
it is found 
that occupied KS orbital energies approximate the experimental IPs of molecules much closer ($\sim$0.1~eV difference) than those of HF 
($\sim$1~eV difference)~\cite{CGB2002,GB2004,BGM2013}. 
Although the question about the physical content of the KS orbitals and the KS orbital energies raised a scientific debate~\cite{PY1989}, theoretical justification for this result 
was given by Baerends and co-workers~\cite{CGB2002,GBB2003} and by Bartlett and co-workers~\cite{BLS2005,VB2012} who proved a generalization of Koopmans' theorem in KS-DFT.
The KS molecular orbitals are routinely employed for chemical applications~\cite{SH1999,KBP1996,BG1997}. 

Orbital energies from local or semilocal density functional approximations (DFAs) underestimate substantially the 
IPs of molecules~\cite{PN1982,PA1998}. 
Nevertheless, they are still useful and the agreement 
with experimental IPs can be improved by applying a uniform shift~\cite{BR1973,PA1998} or linear scaling~\cite{SH1999}. 
The wrong asymptotic behavior of the KS potential, a major deficiency of local DFAs like the local density approximation 
(LDA) or the generalized gradient approximation (GGA) and the manifestation of self-interactions~\cite{P1990} (SI's), is  
responsible for the large deviations of orbital energies from experimental IPs. 
For example, at large distances, the LDA exchange and correlation (xc) potential vanishes exponentially fast, rather than correctly as $-1/r$. 
Consequently, the electron-electron (e-e) part of the approximate KS potential decays as $N/r$ and an electron of the system at infinity, feels the repulsion of all $N$ electrons, itself included. 
The Perdew-Zunger self interaction correction (SIC) method~\cite{PZ1981} in DFT offers a 
correction to this problem and was found to yield orbital energies closer to the
experimental IPs~\cite{GU1997} improving several other properties as well~\cite{GU1997,TS2007,FS2003}. 

The $GW$ method\cite{H1965,HL1986,ORR2002} was initially introduced to improve the obtained quasiparticle spectrum of solids but 
in the last decade, $GW$ at various levels of approximations was also applied to finite 
systems\cite{BAO2011,FAORB2011,MCRHKCRSR2012,CRRSR2012,CRRRS2013,CARRSR2014,SWE2013}
 improving significantly the quasiparticle excitation energies with respect to standard DFT-approximations. 
Those calculations suffer from a strong initial state dependence and the good agreement found could be just fortuitous.  
In this context, self-consistent $GW$\cite{CRRSR2012,CRRRS2013,CARRSR2014} 
 was found to systematically improve ionization energies and total energies of closed shell systems. 
The single electron spectral properties are in very close agreement with experiment avoiding the starting-point dependence.

Recently, Gidopoulos-Lathiotakis~\cite{GL2012} proposed to deal with the problem of SIs in DFAs by 
replacing the approximate Hartree exchange and correlation potential in the KS equations, with a 
different effective potential. The latter is obtained from the optimization of the same DFA energy, but is 
further constrained to satisfy conditions that enforce on it the asymptotic behavior 
of the exact KS potential. The resulting optimal potential was found to improve dramatically 
the agreement of orbital energies with 
the experimental IPs.

Reduced-density-matrix-functional theory (RDMFT) was introduced~\cite{G1975} as an alternative framework to DFT.  
In RDMFT, the one-body reduced density matrix (1-RDM) is the fundamental variable, 
in place of the 
electron density. 
Basic quantities associated with the 1-RDM are the occupation numbers and the natural orbitals, i.e.,\ its eigenvalues and eigenfunctions.
Several approximations for the total energy as a functional of the 1-RDM -- or usually in terms
of the occupation numbers and the natural orbitals --
have become available~\cite{M1984,GU1998,bb0,GPB2005,power_finite,pernal2010,AC3,pade,pnof1,pnof2,PNOF5,sharma08}.
They have proven to describe correctly many diverse properties such as molecular  
dissociation~\cite{bb0,GPB2005,power_finite,pernal2010,AC3} 
or band gaps~\cite{sharma08,helbig09,dfg10,SDSG2013}. However, so far, the 
computational cost has restricted the applications of RDMFT to prototype systems. 
Most of the computational expense is due to the determination of the orbitals
which are not obtained from an eigenvalue equation but through a
numerically expensive minimization.
In contrast to DFT, in RDMFT there is no KS noninteracting system with the same 
1-RDM as the interacting system. Thus different approaches have to be considered in order to define 
effective single-particle Hamiltonians~\cite{pernal_epot,piris_jcc,Baldsiefen2013114}.


Recently, we proposed local-RDMFT~\cite{LHRG2014}, 
a theoretical framework that incorporates static correlation effects in the single-particle, Kohn-Sham equations.
It is based on the adoption of RDMFT approximate functionals (optimized with fractional occupations) 
for the exchange and correlation energy, together with a search for an effective 
local potential, whose eigen-orbitals minimize the total energy.
The search of the effective potential is performed as in Ref.~\cite{GL2012}, 
where, apart from correcting possible SIs, it was also found to avoid mathematical pathologies of finite-basis 
optimized effective potential (OEP)~\cite{GL2012b}. 

Local-RDMFT can be viewed within the framework of the OEP method in DFT, where the correlated exchange and correlation (xc) functionals from RDMFT allow us 
to go beyond the level of an exchange only OEP (x-OEP) calculation~\cite{kummel2008}. 
Equally, local-RDMFT can be regarded an approximation in RDMFT, employing an effective single particle scheme to generate the approximate natural orbitals (ANOs). 
Thus, local RDMFT provides an energy eigenvalue spectrum directly connected to the ANOs and
as we find in Ref.~\cite{LHRG2014}, this energy spectrum reproduces the
 IPs of  small molecules in closer agreement with experiment than HF Koopmans'. 
In addition, it allows us to calculate accurately total energies at any geometry, from equilibrium all the way 
to the dissociation limit, which is well described without the need to break any spin symmetry.  

In the present work, we demonstrate the efficiency of local-RDMFT by applying it to larger molecules.
%
More specifically, we calculate the IPs of  systems 
of $\sim20$ atoms and compare them with experiment. 
For some aromatic molecules we compare the calculated orbital energies with the peaks of the corresponding 
photoelectron spectra (PES). We also study
the relative stability of C$_{20}$ isomers and show that systems of this size are within 
the reach of our method.
Finally, we propose that, in local-RDMFT, 
the optimization of the fractional occupations can be simplified 
by modelling them in terms of the orbital energies. We expect that such ideas will be very useful in the application 
to larger systems and solids.

In Sec.~\ref{sec:theory}, we summarize the basics of local-RDMFT and in Sec.~\ref{sec:appl} we
discuss our results on the application to the C$_{20}$ isomers, the IPs of molecular systems and the 
comparison of the calculated orbital energies with the PES of aromatic molecules.
Finally, in Sec.~\ref{sec:model}, we demonstrate that the optimization of fractional occupation numbers can be
simplified by modeling their dependence on single particle energies. 

\section{Local-RDMFT\label{sec:theory}}

Local-RDMFT combines two main features: (i) the non-idempotency of the optimal 1-RDM 
where the fractional occupation numbers are provided by minimizing the total energy functional 
under Coleman's $N$-representability conditions and (ii) the incorporation of a single 
particle effective Hamiltonian with a local potential. 
As we showed in~\cite{LHRG2014} one has to depart from xc functionals that are explicit functionals 
of the electronic density alone, since
they lead to idempotent solutions. 
Thus, we have to adopt either explicit functionals of the 1-RDM, or functionals of the orbitals and the
occupation numbers.

The central assumption in local-RDMFT~\cite{LHRG2014} is that the search for the set of optimal ANOs 
is restricted in the domain of orbitals that satisfy single-particle equations (KS equations) with 
a local potential. 
The search for the e-e repulsive part $V_{\rm rep}({\br})$ of the effective local potential 
(the analogue of the Hartree-exchange and correlation potential in the KS equations) is 
effected indirectly, by a search for the effective repulsive density (ERD)
$\rho_{\rm rep}({\br})$ whose electrostatic potential is $V_{\rm rep}({\br})$, i.e.,
\begin{equation}
\nabla^2 V_{\rm rep}({\br}) = - 4 \pi \rho_{\rm rep}({\br})\,. 
\end{equation}
Additionally, following Ref.~\cite{GL2012},
 two constraints are imposed in the minimization with respect to $\rho_{\rm rep}({\br})$: 
\begin{eqnarray}
\int d{\br} \: \rho_{\rm rep}({\br}) &=& N-1 \label{eq:asympt}\\
\rho_{\rm rep}({\br}) &\ge& 0\,. \label{eq:pos}
\end{eqnarray}
 The first condition 
is a property of the exact KS potential and the x-OEP potential, 
while the second is a condition that gives physical content to the ERD as a density of $N-1$ electrons.
It is unknown if (\ref{eq:pos}) is a property of the exact KS potential or of x-OEP but without it, 
the search for ERD is mathematically ill posed for finite basis sets. The two conditions, (\ref{eq:asympt}), (\ref{eq:pos}), together 
lead to physical solutions. The optimal ERD and the effective local potential can be obtained, 
similarly to the OEP method, by solving the integral equation~\cite{LHRG2014}
\begin{equation}\label{eq:rho}
\int d^3 r' \, {\tilde \chi} (\v r , \v r') \, \rho_{\rm rep} (\v r') = {\tilde b}( \v r), 
\end{equation}
with
\begin{equation}
\label{eq:oep2}
{\tilde \chi} ( \v r , \v r')\! \doteq\!\! \iint \!\! d^3 x \, d^3 y \, \frac{ \chi ( \v x , \v y) }{ | \v x - \v r| | \v y - \v r'| }\,, 
\end{equation}
\begin{equation}
{\tilde b} ( \v r ) \! \doteq \int \!\!   d^3 x \, \frac{ b ( \v x ) }{ | \v x - \v r|  }\, .
\end{equation}
The response function $\chi (\v r, \v r')$ and $b (\v r)$ are  given by
\begin{equation}
\label{eq:A}
   \chi (\v r , \v r') = {\sum_{j,k,j \ne k}}  \phi_j^*(\v r)\,\phi_k(\v r)\,\phi_k^*(\v r')\,\phi_j(\v r')\frac{n_j-n_k}{\epsilon_j -\epsilon_k}\,, 
\end{equation}
\begin{equation}
   b(\v r) = {\sum_{j,k,j \ne k}} \langle \phi_j | \frac{F_{\rm Hxc}^{(j)}-F_{\rm Hxc}^{(k)}}{\epsilon_j - \epsilon_k} 
| \phi_k \rangle\,  \phi_k^*(\v r)\, \phi_j(\v r)\,,
\label{eq:B}
\end{equation}
with $F_{\rm Hxc}^{(j)}$ defined by 
\begin{equation}
\label{eq:Liapis}
\frac{\delta E_{\rm Hxc}}{\delta \phi_j^*(\v r)} \doteq
\int d^3r' \, F_{\rm Hxc}^{(j)}(\v r, \v r') \, \phi_j(\v r')\,. 
\end{equation}
$E_{\rm Hxc}$ is the approximation for the e-e 
interaction 
energy,  $\phi_j$ are the ANOs and $n_j$, $\epsilon_j$ their corresponding occupation numbers
and orbital energies (eigenvalues of the effective hamiltonian).
The two constraints can be incorporated with a 
Lagrange multiplier (\ref{eq:asympt}) and a penalty term (\ref{eq:pos}) that introduces an energy cost 
for every point $\br$ where $ \rho_{\rm rep}(\br)$ becomes negative.

Terms over pairs of orbitals with almost equal occupations cause numerical instabilities in the sums 
of Eqs.~(\ref{eq:A}) and  (\ref{eq:B}) and we have decided to exclude them by introducing a small cutoff 
$\Delta n_c$. The reader is referred to the discussion in Ref.~\cite{LHRG2014}. Our choice affects mostly 
pairs of weakly occupied orbitals, whose energies are in any case inaccurate for finite localized orbital 
basis sets, as occasionally they violate the aufbau principle and the negative definiteness of $\chi$. 
For very small cutoff $\Delta n_c$ we observe convergence problems, mainly while attempting to enforce the 
positivity constraint (\ref{eq:pos}). When $\Delta n_c$ is large enough  to exclude erroneous terms
involving weakly occupied orbitals (past a typical value $\sim 0.1$), convergence issues improve dramatically 
and IPs remain unchanged for a broad range of values of $\Delta n_c$.
We have found that a choice for  $\Delta n_c \sim 0.1-0.3$ leads to sTable solutions where the IPs are insensitive to a change of $\Delta n_c$.

The ANOs are expanded in a basis
set (orbital basis) while the ERD in a separate (auxiliary) basis and Eq.~(\ref{eq:rho}) 
transforms into a linear system of equations. This linear system typically becomes singular
and we use the singular value decomposition (SVD) to obtain smooth 
and physical densities and potentials (see Refs.~\cite{GL2012} and \cite{LHRG2014}). 
We note that a SVD for the matrix of the density-density response function may introduce singular behavior
in the effective potential \cite{GL2012b}. However, the 
two constraints (\ref{eq:asympt}), (\ref{eq:pos}) reduce the variational freedom in the 
space of effective local potentials $V_{\rm rep} (\br)$, to such a degree that a discontinuity 
correction~\cite{GL2012b} in the null space of the (finite-orbital-basis) response function is no longer necessary. 

We stress that there is no functional derivative relation linking our local effective potential
with the total energy functional. As a result, in our formulation, we avoid the collapse of  all eigenvalues 
corresponding to fractional occupations to the chemical potential. 
In addition, 
the effective local potential in local-RDMFT differs in general from the exact KS potential.  
However, a comparison with the exact KS potential is still meaningful and establishes the physical significance of the approximation. 

\section{Applications\label{sec:appl}}

\begin{table*}
\begin{center}
\def\arraystretch{0.85}
\begin{tabular}{l|ccccccccc}
System		& HF Koopmans' &  Mueller      & BBC3         &  Power       & ML           &  Exp.\\
\hline\hline
Benzene 	& 9.07(-1.69)  &  9.65(4.53)   & 8.95(-3.03)  &  9.42(2.08)  & 9.30(0.81)   & 9.23 \\
Pyridine	& 9.33(0.78)   &  9.77(5.51)   & 8.86(-4.32)  &  9.62(3.89)  & 9.55(3.13)   & 9.26\\
Naphthalene	& 7.80(-3.54)  &  8.26(2.13)   & 7.54(-6.84)  &  7.77(-4.01) & 7.84(-3.04)  & 8.09\\
Phenanthrene	& 7.62(-3.71)  &  7.58(-4.17)  & 6.83(-13.65) & 7.03(-11.13)& 7.10(-10.24) & 7.91\\
Anthracene	& 6.91(-6.95)  &  7.32(-1.48)  & 6.37(-14.27) & 6.74(-9.29) & 6.85(-7.81)  & 7.43\\
Pyrene		& 6.97(-6.05)  &  7.24(-2.43)  & 6.31(-14.96) & 6.63(-10.65)& 6.64(-10.51) & 7.42\\
Methane		& 14.77(8.57)  &  13.69(0.66)  & 13.51(-0.66) & 13.53(-0.51)& 13.93(2.43)  & 13.60,14.40\\
Ethane		& 13.13(9.53)  &  11.81(-1.50) & 12.37(3.17)  & 12.02(0.25) & 12.62(5.25)  & 11.99\\
Propane		& 12.63(9.77)  &  11.68(1.48)  & 11.50(-0.09) & 11.62(0.96) & 12.15(5.56)  & 11.51\\
Butane		& 12.37(11.54) &  11.32(2.07)  & 11.18(0.81)  & 11.33(2.16) & 11.80(6.40)  & 11.09\\
Pentane		& 12.14(11.39) &  10.89(-0.09) & 10.76(-1.28) & 10.05(-7.80)& 11.47(5.23)  & 10.90\\
Cyclo-Pentane	& 12.14(10.29) &  11.20(1.73)  & 11.25(2.18)  & 11.25(2.18) & 11.75(6.72)  & 11.01\\
Hexane		& 11.93(17.73) &  10.64(5.03)  & 10.57(4.34)  & 10.77(6.32) & 11.14(9.97)  & 10.13\\
Cyclo-Hexane-b	& 11.52(11.61) &  10.72(3.88)  & 10.86(5.23)  & 10.72(3.88) & 11.16(8.14)  & 10.32\\
Cyclo-Hexane-c	& 11.52(11.62) &  10.82(4.84)  & 10.83(4.94)  & 10.93(5.91) & 11.08(7.36)  & 10.32\\
Heptane		& 11.77(18.50) &  10.30(3.73)  & 10.25(3.22)  & 10.50(5.74) & 10.89(9.67)  & 9.93\\
Octane		& 11.64(18.80) &  10.20(4.08)  & 9.99(1.94)   & 10.28(4.90) & 10.66(8.78)  & 9.80\\
Methanol	& 12.05(9.91)  &  10.40(-5.11) & 9.90(-9.67)  & 10.39(-5.20)& 11.20(2.19)  & 10.96\\
Ethanol		& 11.84(11.28) &  10.39(2.35)  & 9.35(-12.12) & 10.24(3.76) & 10.97(3.10)  & 10.64\\
Propanol	& 11.83(12.59) &  10.17(-3.24) & 9.32(-11.32) & 10.28(-2.19)& 11.00(4.66)  & 10.51\\
Azulene		& 6.99(-5.80)  &  7.72(4.04)   & 7.08(-4.58)  & 7.29(-1.75) & 7.27(-2.02)  & 7.42\\
Ethylene	& 11.11(4.03)  &  10.76(0.75)  & 10.21(-4.40) & 10.39(-2.72)& 10.46(-2.06) & 10.68\\
Butadiene	& 8.66(-4.13)  &  8.90(-1.44)  & 8.37(-7.31)  & 8.77(-2.88) & 8.78(-2.77)  & 9.03\\
Hexatriene	& 7.87(-5.20)  &  8.02(-3.37)  & 7.27(-12.41) & 7.84(-5.54) & 7.73(-6.87)  & 8.30\\
Octatetraene	& 7.37(-5.37)  &  7.46(-4.24)  & 6.65(-14.63) & 7.24(-7.06) & 7.17(-7.96)  & 7.79\\ \hline
$ \Delta $	& {\bf 8.81}   &  {\bf 2.96}   & {\bf 6.46}   & {\bf 4.51}  & {\bf 5.49}   &     \\
$ \sigma $      & {\bf 1.04}   &  {\bf 0.32}   & {\bf 0.71}   & {\bf 0.48}  & {\bf 0.61}   &     \\
\hline
\end{tabular}
\end{center}
\caption{ \label{tab:IPbig} IPs (in eV) for several molecules obtained as the HOMO energy of the effective Hamiltonian employing
several RDMFT functionals and compared with HF Koopmans' (with the same basis set) and experiment.
The percentage errors compared to experiment are in parenthesis. The average absolute percentage error,
$\Delta =100\times(1/N)\sum_i |(x_i - x_i^{\rm ref})/x_i^{\rm ref}|$, 
and the root-mean-square deviation $\sigma = (1/\sqrt{N}) \sqrt{\sum_i (x_i - x_i^{\rm ref})^2}$
are also included. 
The experimental IPs in the last column are
obtained from NIST Chemistry WebBook~\cite{NIST}. Vertical IP values are preferred when available.}
\end{table*}

\begin{figure} 
\begin{center}
\includegraphics[width=7cm, clip]{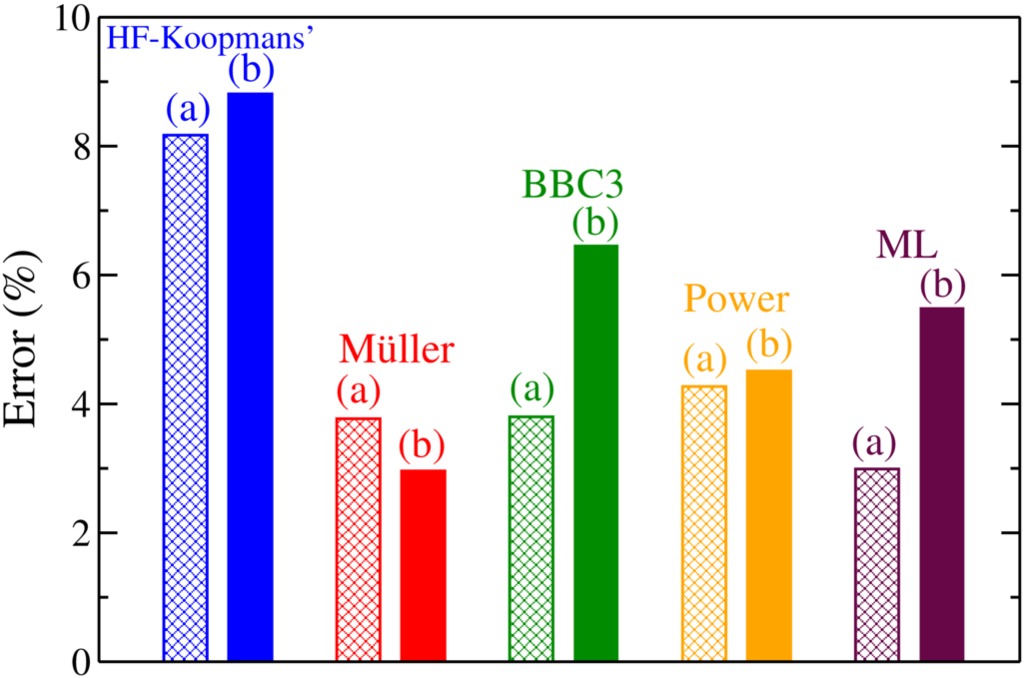}  
\end{center}
\caption{\label{fig:ips} Average, percentage, absolute errors in the IPs calculated as the orbital energies
of the local-RDMFT effective Hamiltonian for several RDMFT approximations for (a) the small atoms and molecules
shown in detail in Ref.~\cite{LHRG2014} and (b) those in Table~\ref{tab:IPbig} compared with experimental results.}
\end{figure}

We applied local-RDMFT  
to the molecular systems shown in Table~\ref{tab:IPbig}, employing several approximate RDMFT  functionals:
M\"uller~\cite{M1984,bb0}, the third Buijse-Baerends corrected (BBC3) approximation~\cite{GPB2005},
the Power functional~\cite{sharma08,power_finite} and the Marques-Lathiotakis (ML) approximation~\cite{pade}. 
Values obtained with HF Koopmans' are also included.
The cc-pVDZ and the uncontracted cc-pVDZ basis sets were employed for the orbital and the auxiliary 
basis sets respectively for all calculations. 
 Our target is to examine
the usefulness of the quasiparticle energy spectrum, i.e.,\ how close the orbital energies are
(in absolute value) to the corresponding IP.
Only the first IPs are shown in Table~\ref{tab:IPbig} for simplicity.
Such a comparison for small atomic and molecular systems 
(He, H$_2$, Be, Ne, H$_2$O, NH$_3$, CH$_4$, CO$_2$, C$_2$H$_2$, C$_2$H$_4$) 
is shown in Ref.~\cite{LHRG2014} using larger basis sets and for up to the three 
IPs for the same molecule. 
In Fig.~\ref{fig:ips}, the present results for larger systems are compared to those in Ref.~\cite{LHRG2014} by plotting the absolute, percentage error of the IPs for the four different 
RDMFT energy functionals and HF.
We find a remarkable agreement between the energy eigenvalues and the
experimental IPs for the functionals we tested.
For the small systems, all errors are below or around 5\%, with the ML functional 
as low as $\sim$3\%. We should emphasize that this agreement is found only if the 
positivity condition of Eq.~(\ref{eq:pos})
 is enforced, otherwise the error increases to 20\% for some functionals. 
The agreement with experiment is even better for the first IPs for small systems
with the M\"uller functional the most accurate in this case with an error of only $\sim$2\%. 
The trend is similar for the larger systems of Table~\ref{tab:IPbig} where the average error is relatively larger 
for all functionals except for the M\"uller functional, for which it remains below 3\%. 
Overall the agreement with experimental values is very good and substantially 
better than HF Koopmans'. 
The good performance of the M\"uller functional is rather uniform. 
In contrast, the results of BBC3  deviate from experiment mainly for the IPs of alcanes, raising 
the error to $\sim$6.5\%.

\begin{figure}
\vspace{0.4cm}
\begin{center}
\begin{tabular}{c}
\includegraphics[width=7cm]{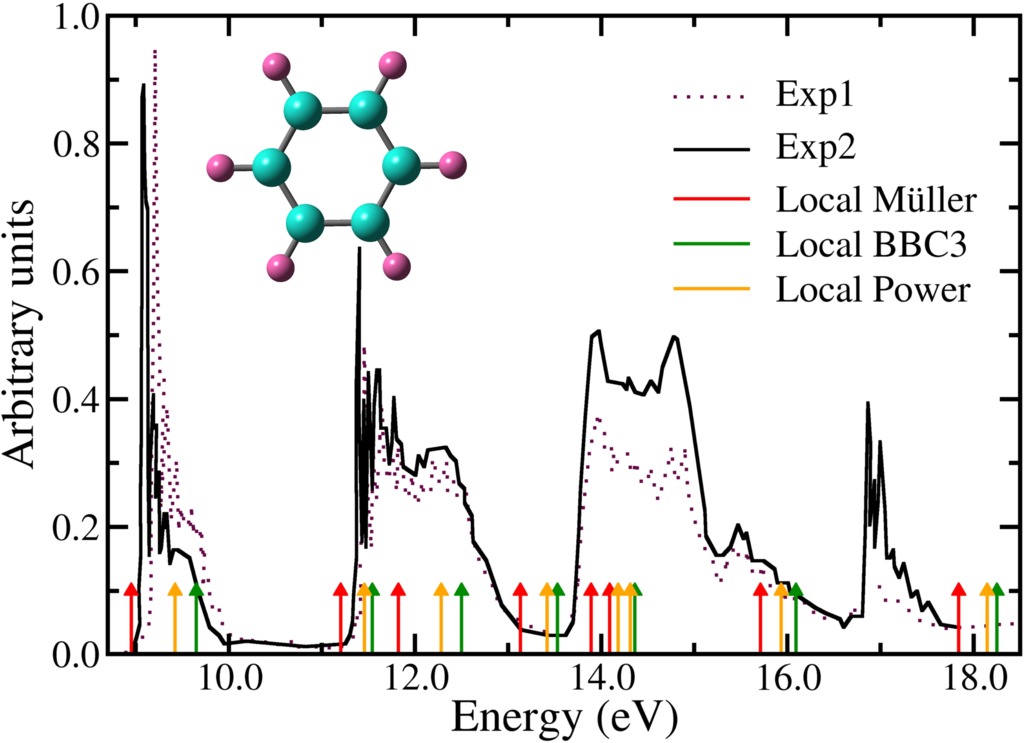} \\
\includegraphics[width=7cm]{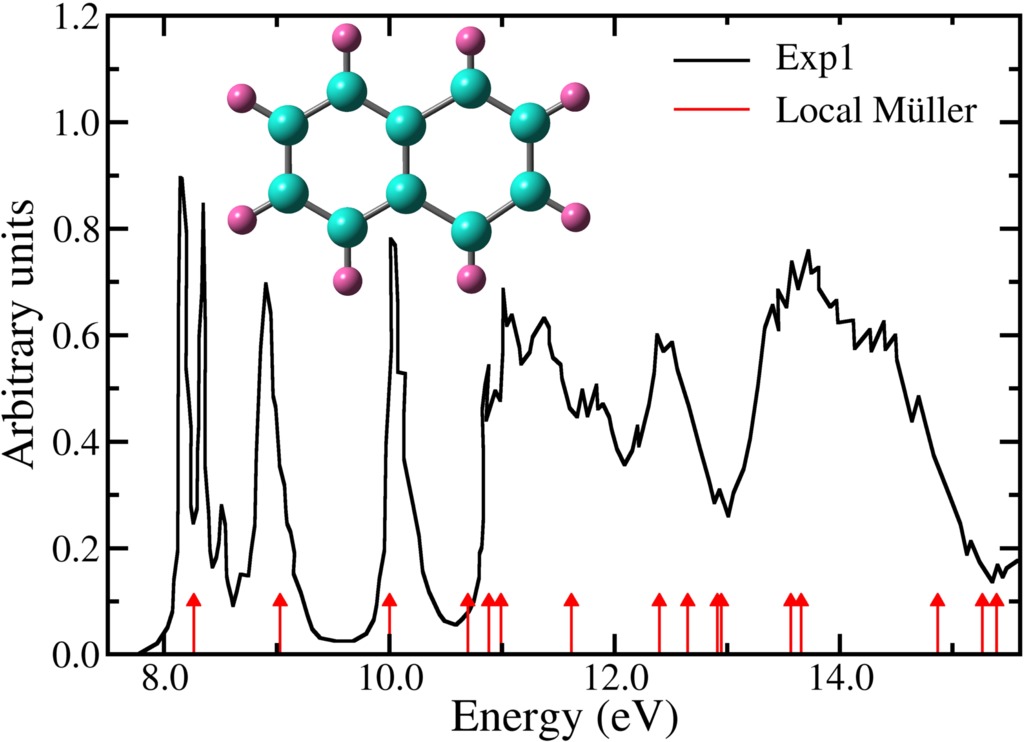} \\
\includegraphics[width=7cm]{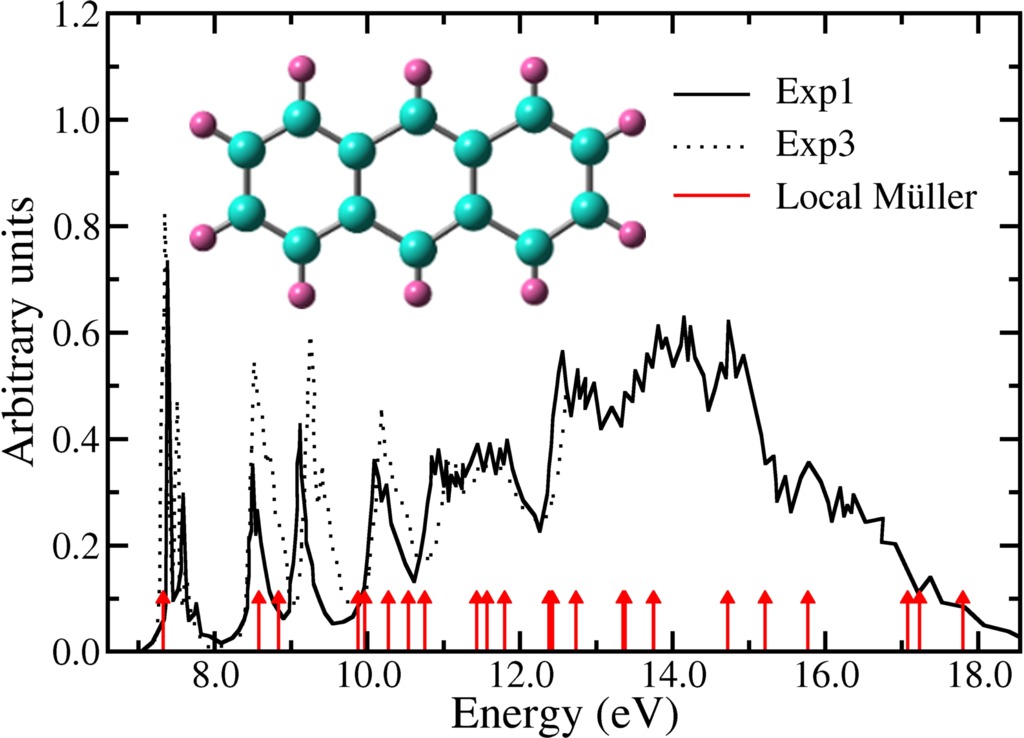} \\
\end{tabular}
\end{center}
\caption{\label{fig:aromatics}
Comparison of the local-RDMFT eigenvalues (vertical lines)  
with measured PES for benzene (top), naphthalene (middle) and
anthracene (bottom)
(Exp1: Ref.~\cite{CBH1972}, Exp2: Ref.~\cite{SK1978}, Exp3: Ref.~\cite{Streets197471}).} 
\end{figure}

As an additional test for the physical interpretation of the obtained quasiparticle energy spectrum, we 
plot (see Fig.~\ref{fig:aromatics}) the orbital energies of local-RDMFT together with the experimental 
PES for three aromatic molecules, benzene, naphthalene and anthracene. For all 
systems, we show the eigenvalues obtained with the M\"uller functional as this functional is found 
to yield better results. For benzene, we also include the BBC3 and power functional eigenvalues. The 
agreement with the experimental spectrum is fair for all three functionals, however, the M\"uller 
orbital energies are more accurate. Especially for naphthalene, the M\"uller eigenvalues are in
excellent agreement with experiment.

For comparison, for the LDA and GGA approximations, the errors of the KS eigenvalues compared with experimental IP's 
 can be of the order of 30-40\% due to
SIs\cite{GL2012,SWE2013}. The inclusion of a percentage of HF exchange in hybrid functionals reduces 
SIs which, however, remain large. As an example, we applied the Becke 3 parameter exchange-correlation
functional\cite{B1993} (B3LYP) to the systems in Table~\ref{tab:IPbig}
(using the same geometries and basis set) and the absolute, percentage error of the IPs is 26\%.

$GW$ method applied on top of plain DFT and HF calculations is found to improve substantially the quasiparticle
spectrum. To summarize a few applications: Blase et al\cite{BAO2011} obtained the IPs of photovoltaic-relevant molecules
with a mean average error of 3.8~\%. Van Setten et al\cite{SWE2013} found the IPs of 27 molecular systems with a 
root-mean-square deviation of 0.47 eV from experiment. Marom et al \cite{MCRHKCRSR2012} assessed the performance of a
hierarchy of $GW$ approximations for benzene, pyridine, and the diazines and compared the quasiparticle spectrum 
with PES. Caruso and co-workers\cite{CRRSR2012,CRRRS2013,CARRSR2014} developed an all electron implementation of self consistent $GW$ (sc-$GW$) with localized basis functions
and showed that it is more accurate than other approximations lower in $GW$ hierarchy. They applied sc-$GW$ on five molecules 
relevant for organic photovoltaics\cite{CRRRS2013} obtaining an average error of 0.4 eV (maximum error 1.2 eV). 
%

The $GW$ calculations come with a high computational cost
despite employing routinely efficiency improving techniques like the resolution of 
identity\cite{1367-2630-14-5-053020}. Local-RDMFT on the other hand is more efficient
method and the efficiency can further improve by adopting techniques like the RI. More importantly, it provides IPs of similar 
quality as $GW$ approaches. Although we cannot make a quantitative comparison since we used different set of systems and
basis sets, as we see in Table~\ref{tab:IPbig}, the root-mean-square deviation from experiment
for the local M\"uller functional is as low as 0.32. This quantity has values similar to those reported for the $GW$ approaches for all
the functionals we employed.

As a demonstration of the efficiency of local-RDMFT we calculated the three most stable isomers
of C$_{20}$ namely the cage, the bowl and the monocyclic ring structures using several functionals. 
These three isomers are energetically very close and the predicted most stable isomer 
differs from method to method. For example, the ring is found the most stable by HF~\cite{FGSA1992}
and DFT-GGA~\cite{RSOSPJG1993} and B3LYP~\cite{XZZCZS2006},
the cage by DFT-LDA~\cite{Wang1996121} and CCSD~\cite{Taylor1995558} 
and the bowl by a more recent CCSD~\cite{AGBZ2005} and quantum Monte Carlo calculations~\cite{Sokolova2000229}. 
CCSD and 
QMC are the most accurate schemes, and there is a consensus that the bowl is the most stable
structure with the cage being almost isoenergetic. Our local-RDMFT results for the total energies of the
three isomers are shown in Table~\ref{tab:c20}. We employed cc-pVDZ and uncontracted 
cc-pVDZ as orbital and auxiliary basis sets, respectively, and the optimal geometries that were 
obtained at the 
MP2 level of theory. Apart from the approximations considered above in this application we also employed 
the functionals of Goedecker-Umrigar (GU)~\cite{GU1998}, the automatic third correction (AC3)~\cite{AC3} 
and the first Piris natural orbital functional (PNOF1)~\cite{pnof1}.
In agreement with other methods, we find that with all functionals the 3 isomers are close in energy,
especially the bowl and the cage. 
Most functionals predict the cage to be slightly more stable 
while only PNOF1 predicts the bowl. Calculations with MP2 and MP4 theories~\cite{refdata,g09}
with the same basis set and geometries also show the cage structure to be the most stable. 
Hence, while our results are not in agreement with QMC, it is not conclusive if the difference is due to the employed method 
or the numerical details of the calculation. However, the purpose of this application is to demonstrate that problems 
of this scale are tractable with local-RDMFT yielding sensible results. Probably, more sophisticated functionals within RDMFT
are required to capture the delicate energy differences of C$_{20}$ isomers more accurately. 
Finally, the IPs of the 
C$_{20}$ isomers calculated as the energy eigenvalue of the HOMO using the M\"uller
functional are 7.1, 8.6, and 7.0 eV for the ring, the bowl and the cage isomers, respectively. 
These values are
in very good agreement with the IPs obtained by total energy difference 
at the MP4 level of theory~\cite{refdata,g09}, 7.26, 8.92, and 6.98~eV, respectively. 

\begin{table}
\caption{\label{tab:c20}
Total energies (in a.u.) for the three most stable C$_{20}$ isomers obtained 
with various local-RDMFT functionals and MP2, MP4 theories.
The most stable structure for a given approximation is given in bold face. 
}
\begin{tabular*}{\hsize}{@{\extracolsep{\fill}}lrrr}
     & Ring & Bowl & Cage\\ \hline
M\"uller  & -761.33 & -761.37 & {\bf -761.38}\\
BBC3  & -758.47 & -758.44 & {\bf -758.55}\\
Power & {\bf -758.93} & -758.83 & -758.85\\
ML   & -758.23 & -758.26 & {\bf -758.30}\\
GU   & {\bf -760.59} & -760.29 & -760.34\\
AC3  & -758.58 & -758.55 & {\bf -758.64}\\
PNOF1  & -756.62 & {\bf -756.65} & -756.54\\ \hline
MP2  & -759.23 & -759.32 & {\bf -759.34} \\
MP4  & -759.40 & -759.47 & {\bf -759.51} \\
\end{tabular*}
\end{table}

The results presented in this section demonstrate that the local-RDMFT formalism preserves the
advantages of RDMFT in calculating correlation energies and, as we showed in Ref.~\cite{LHRG2014}, 
also in describing molecular
dissociation. In addition, it provides energy eigenvalues which are in very good
agreement with experimental IPs. Compared with standard RDMFT, the significant reduction in
computational cost allows for applications to larger systems previously
inaccessible to this theory.

\section{Modelling the fractional occupancies\label{sec:model}}

 Fractional occupation numbers are usually employed in DFT calculations in an ad-hoc way to introduce 
temperature effects and to help the convergence of the self-consistent KS-equations loop in small-gap or metallic 
systems. 
In the case of local-RDMFT, fractional occupations are introduced naturally through an optimization
procedure. 
The existence of fractional occupations and at the same time of a corresponding single electron energy spectrum allows for 
the modelling of the occupation numbers as functions of the energy eigenvalues. 
This modelling is no longer arbitrary and is implemented through the functional optimization. 
For instance, one can assume a smooth parametric form for the function $n_j (\epsilon_j)$  connecting the 
occupation numbers to single-particle energies and optimize the model parameters such that 
the energy functional is minimized.   
The advantage is that occupation numbers are obtained in a simpler minimization procedure of a few variables only.
As a demonstration we consider
\begin{equation}
n_j(\epsilon_j) = \frac{1}{1+e^{\beta (\epsilon_j-\mu)}}, \quad \beta=\left\{ 
\begin{array}{l} \beta_s, \ \ \ \ \epsilon_j < \mu \\ \beta_w, \ \ \ \  \epsilon_j > \mu\,. \end{array}
\right.
\label{eq:fermi}
\end{equation} 
The parameters $\mu$, $\beta_s$ and  $\beta_w$ are optimized by minimizing the energy functional with respect to them
for a given set of orbitals. 
A Fermi distribution modelling of the occupation numbers as functions of the eigenvalues of a Hamiltonian with
a local potential was also introduced by Gr\"uning et al.~\cite{GGB2003}. However, in that work, the model parameters
were not optimized iteratively with the orbitals for each calculation but were chosen universally such that the obtained
dissociation of H$_2$ molecule is as close to the exact as possible. 

\begin{figure}
\vspace{0.4cm}
\begin{center}
\includegraphics[width=8cm]{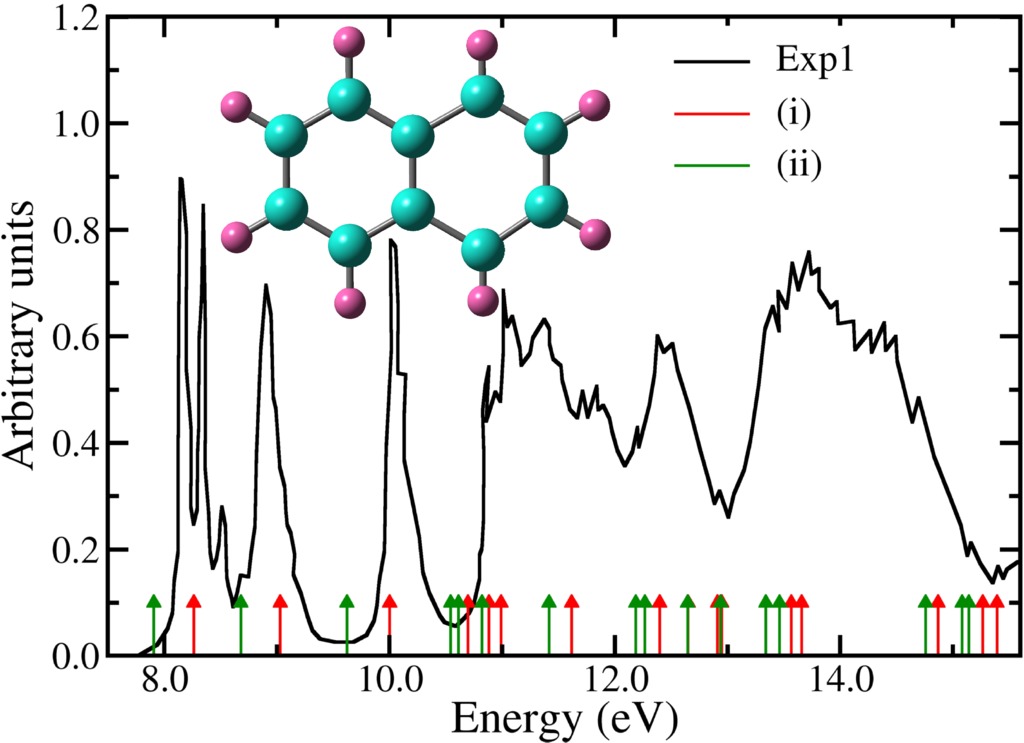}
\end{center}
\caption{\label{fig:Fermi}
Comparison of the local-RDMFT eigenvalues (vertical lines) using the M\"uller functional, 
with the experimental PES, Exp1:~\cite{CBH1972}, for naphthalene using 
(i) local-RDMFT with full occupation number optimization and (ii) through the optimization of the parameters of the function $n_j(\epsilon_j)$ introduced in Eq.~(\ref{eq:fermi}). 
}
\end{figure}

In Fig.~\ref{fig:Fermi}, we show the energy eigenvalues obtained for naphthalene compared with the 
PES and the eigenvalues of the standard local-RDMFT. 
As we see the obtained spectrum of such a model is reasonable.  
We believe that such a procedure will
be useful in the application to periodic systems, especially metals, 
simplifying the optimization of the occupation numbers, offering a natural way to introduce occupation 
smearing quasiparticle renormalization factors 
and accounting for quasiparticle renormalization effects in the homogeneous 
electron gas case~\cite{LHG2007}.

\section{Conclusion and perspectives}

Local-RDMFT, a novel scheme that incorporates static correlation in the KS equations and allows the 
accurate description of molecular dissociation, was applied to molecular systems of size up to 20 atoms.

The new approach associates a quasiparticle energy spectrum to
the ANOs. This spectrum is in good agreement with experimental IPs and 
PES for molecules. The reduction in computational cost, permitted, 
for the first time, the calculation of larger molecules with the improved accuracy of RDMFT 
functionals. To demonstrate the efficiency of the new scheme, we applied it to 
the three most stable C$_{20}$ isomers although the tiny energy differences of these 
systems are probably beyond the accuracy of current RDMFT approximations.

The new method provides a powerful tool which opens a new avenue for bringing the advantages of 
RDMFT into DFT. Due to the similarity of the local-RDMFT and the OEP 
equations, the systematic and physically motivated approximations in density-matrix based 
schemes to cope with strongly correlated systems~\cite{sharma08} and static correlation can 
now easily be brought to the realm of DFT. For the first time, a method is able to 
simultaneously describe ground-state properties, bond-breaking and photoelectron spectra.

Compared with orbital-dependent functionals in DFT, the additional cost in local-RDMFT 
comes from the iterative optimization of the occupation numbers and the ANOs. This extra 
cost can be reduced by connecting the occupation numbers directly to the energy eigenvalues 
through physically motivated models, see equation~(\ref{eq:fermi}). 

In the future, the method can be extended to the time-dependent regime with the aim to 
provide more accurate energy spectra and description of electronic excitations. The development 
of a linear-response formalism will in addition give access to a large number of experimentally 
measurable properties. 

\begin{acknowledgments} 
NNL acknowledges financial support from the GSRT action $\rm KPH\Pi I\Sigma$,
project ``New multifunctional Nanostructured Materials and Devices - POLYNANO'', No. 447963,
NH from a DFG Emmy-Noether grant,
and AR from the European Research Council Advanced Grant (ERC-2010-AdG-267374) 
Spanish Grant (FIS2010-21282-C02-01), Grupo Consolidado UPV/EHU (IT578-13), 
and European Commission project CRONOS(280879- 2). 
\end{acknowledgments}


%

\end{document}